\documentstyle[12pt]{article}
\begin{document}
\baselineskip 24pt
\bibliographystyle{unstr}
quant-ph/9501003
\vbox{\vspace{6mm}}
\begin{center}
{\large \bf \baselineskip=24pt Nonlocality of a Single Photon \break Revisited
Again}
 \\[7mm] Lev Vaidman\\
{\it School of Physics and Astronomy
\\Raymond and Beverly Sackler Faculty of Exact Sciences
\\ Tel--Aviv University,
\ \ Tel--Aviv \ \ 69978 \ \ ISRAEL}\\[5mm]
\end{center}
\vspace{2mm}

Recently, Hardy\cite{H1} argued that the nonlocality of the quantum theory can
be demonstrated for
a single particle. The nonlocality means   the impossibility of
constructing  a local hidden variable
theory reproducing the predictions of quantum theory. However,
Bohm\cite{Bo}  had constructed  such a theory, i.e., hidden variable theory
local at the one
particle level,  and
therefore, Hardy's claim cannot be true. (Bohm's theory is, however,  nonlocal
when
applied to systems consisting of more than just one particle.)

Hardy proposed an experimental setup and correctly analyzed the possible
outcomes of the experiment. However, I believe that  its interpretation  as a
single
photon  experiment is misleading.

 In the usual setup of Bell type
experiments\cite{Be} we have  few systems at  separated locations, one
system at each location.   Hardy's setup does not readily falls into this
category, but if it is, the number of involved quantum systems clearly
larger than one.  Indeed, he has  three
 input channels $s$, $a_1$, and $a_2$ and essentially two separate
locations in which the   clicks of  six  detectors  exhibit  quantum
(nonlocal) correlations.
There is yet another sense of a single particle experiment (which is
probably  closer to  Einstein's vision quoted by Hardy). In this setup
there is a single non-relativistic particle (which cannot be annihilated or
created) with its
Schr\"odinger wave spreaded in space. Obviously,  Hardy's experiment does
not belong to this category either.

If we do allow creation and annihilation of photons, then nonlocality can be
demonstrated using a single photon state, $|\Psi \rangle = \alpha |A\rangle +
\beta
|B\rangle$,  which  is  a superposition of  two
separate wavepackets localized at $A$ and $B$. Aharonov\cite{A}  pointed out
that there is an
isomorphism between  states of this type  and  states of two separate
spin-1/2 particles:
$|\Phi \rangle = \alpha |{\uparrow}\rangle_A  |{\downarrow}\rangle_B +
\beta |{\downarrow}\rangle_A |{\uparrow}\rangle_B$ for which nonlocality is
well established\cite{Be}. The isomorphism alluded above can be realized  by a
physical mechanism which creates locally a photon   when the spin  is ``up'',
and absorbs
a photon when the spin is down:
 \begin{equation}
(\alpha |A\rangle + \beta |B\rangle) |{\downarrow}\rangle_A
|{\downarrow}\rangle_B \leftrightarrow \alpha  |{\uparrow}\rangle_A
|{\downarrow}\rangle_B + \beta  |{\downarrow}\rangle_A
|{\uparrow}\rangle_B.
 \end{equation}
In fact, this Hardy's work is, essentially, a translation of his other result
on
nonlocality for two particles without inequalities\cite {H2}.

Hardy proceeds by presenting a ``paradox''.  He considers  his experiment
in which the outcome was  $F_1=1$ and $F_2=1$. He then points out that in this
case
the photon from the input $s$ invariably has to be found in $u_1$ (if it were
searched
there by detector $U_1$) and, also,   invariably has to be found in $u_2$ (if
it were searched,
instead, by detector $U_2$). He considers this as a paradox since in the
input $s$ we had at most one photon. Hardy  resolves the paradox  by
introducing
 a genuine nonlocality. He claims that placing  detector $U_1$ might
influence the outcome of the measurement in the remote location and we
might not get $F_2=1$. However, there is no reason for his unusual
proposal, since there is no real  paradox to resolve. The correct statement
is instead that the photon invariably
has to be found in $u_1$ if it was  searched by $U_1$ {\it and was not
searched by} $U_2$. Similarly, the photon invariably
has to be found in $u_2$ if it was  searched by $U_2$ {\it and was not
searched by} $U_1$. Clearly, there cannot be a contradiction between these two
correct statements.

Hardy considers here a  pre- and post-selected system  and the feature he
points out is typical for such systems. Probably, the simplest example of
this kind\cite{AAD} is a $single$ particle prepared in a superposition of being
in three boxes
$A, B$ and $C$: $|\Psi_1 \rangle = 1/\sqrt 3 (|A\rangle + |B\rangle +
|C\rangle)$ which was found later in the state $|\Psi_2 \rangle = 1/\sqrt 3
(|A\rangle + |B\rangle -|C\rangle)$. If, in the intermediate time it was
searched in  box $A$ it has to be found there, and if, instead,  it was
searched in  box $B$,  it has to be found there too. (Indeed, not finding
the particle in box $A$ would project the initial state $|\Psi_1 \rangle$ onto
$ 1/\sqrt 2 ( |B\rangle +
|C\rangle)$ which is orthogonal to the final state $|\Psi_2$.) In fact, Hardy
has
previously considered\cite{H3} another, truly  surprising example of this
kind, see
Ref. 8 for our analysis of this example.


\begin{thebibliography} {9}


\bibitem{H1} L. Hardy, Phys. Rev. Lett. {\bf 73}, 2279 (1994)

\bibitem{Bo} D. Bohm, Phys. Rev. {\bf 85}, 180 (1952).

 \bibitem{Be} J.S. Bell, Physics {\bf 1} 195 (1964).

\bibitem{A} Y. Aharonov, Lecture in EPR conference, Switzerland (1985).

\bibitem{H2} L. Hardy, Phys. Rev. Lett. {\bf 71}, 1665 (1993).

\bibitem{AAD}  D. Albert, Y. Aharonov, S. D'Amato, Phys. Rev. Lett. {\bf 54}, 5
(1985).

\bibitem{H3} L. Hardy, Phys. Rev. Lett. {\bf 68}, 2981 (1992).

\bibitem{V} L. Vaidman,  Phys. Rev. Lett. {\bf 70}, 3369 (1993).
\end{thebibliography}
\end{document}